\documentclass{pasa}%

\usepackage{graphicx}
\usepackage{hyperref}
\usepackage{xcolor}

\title[SETI]{A Search for Technosignatures toward the Galactic Centre at 150 MHz}

\author[Tremblay et al.]{Chenoa D. Tremblay$^{1,2}$ \thanks{astrochenoa@gmail.com}, Danny C. Price$^3$, and Steven J. Tingay$^3$ 

\affil{$^1$CSIRO, Space and Astronomy, Australian Telescope National Facility, PO Box 1130, Bentley WA 6102, Australia}
\affil{$^2$SETI Institute, Mountain View, CA 94043, USA }%
\affil{$^3$International Centre for Radio Astronomy Research, Curtin University, Bentley, WA 6102, Australia}}%

\definecolor{dcol}{rgb}{0.7, 0.1, 0.1}

\jid{PASA}
\doi{10.1017/pas.\the\year.xxx}
\jyear{\the\year}

\usepackage{aas_macros}
\usepackage{hyperref} 
\hypersetup{colorlinks,citecolor=blue,linkcolor=blue,urlcolor=blue}

\begin{document}

\begin{frontmatter}
\maketitle

\begin{abstract}
This paper is the fourth in a series of low-frequency searches for technosignatures.  Using the Murchison Widefield Array over two nights we integrate 7\,hours of data toward the Galactic Centre (centred on the position of Sagittarius A$^{*}$) with a total field-of-view of 200\,deg$^{2}$. We present a targeted search toward 144 exoplanetary systems, at our best yet angular resolution (75\,arc\,seconds).  This is the first technosignature search at a frequency of 155\,MHz toward the Galactic Centre (our previous central frequencies have been lower).  A blind search toward in excess of 3\, million stars toward the Galactic Centre and Galactic bulge is also completed, placing an  equivalent isotropic power limit <1.1$\times$10$^{19}$\,W at the distance to the Galactic Centre.  No plausible technosignatures are detected.   
\end{abstract}

\end{frontmatter}

\section{INTRODUCTION }
\label{sec:intro}

The prevalence of life beyond Earth is a central and unanswered question within astrobiology. The search for extraterrestrial intelligence (SETI) seeks to answer this question via detection of ``technosignatures'', artificial signals that indicate the existence of technologically-capable societies \citep[see review by][]{Tarter:2001}. On Earth, low-frequency radio signals, like those used by FM radio, are a ubiquitous choice for communications. Many astrophysical processes give rise to low-frequency radio emission, and as such numerous large and sensitive low-frequency radio telescopes have been built, including the current-generation Murchison Widefield Array \citep[MWA,][]{Tingay_2013, Wayth_PhaseII}, Long Wavelength Array \citep[LWA,][]{Ellingson:2009}, Low-Frequency Array \citep[LOFAR,][]{vanHaarlem:2013} and Giant Metrewave Radio Telescope \citep[GMRT, ][]{Gupta:2017}. The existence of both powerful transmitters and sensitive receivers at low frequencies---both of which emerged early in the history of radio engineering---motivates low-frequency technosignature searches by providing an example class of engineered signals to search for, and instruments with which to do so.

This paper is the fourth in a series of papers detailing SETI observations with the Murchison Widefield Array \citep[MWA,][]{Tingay_2013, Wayth_PhaseII}, the details of which are summarised in Table 1.  The MWA offers two advantages over other SETI searches; its large field of view and the low frequency range. These searches of $\sim$400--600 square degrees, are some of the largest published surveys, although no candidate technosignature signals were detected above the detection limits. Both \cite{Garrett_2017} and more recently \cite{Houston_2021} have discussed the benefits of using aperture arrays like MWA for efficiently completing an all-sky SETI survey. \cite{Houston_2021} outlines strategies of SETI searches from past, present, and future and suggests that if a receiver and transmitter are aligned in ``...space, time and frequency, with adequate receive power, a detection can occur.''  They suggest that unless there is a compelling reason to only search stellar regions, wide-field searches of any signal of unknown origin are required.

However, before we get to all-sky technosignature searches there are a number of computational challenges to overcome and these surveys have provided insight on how to accomplish this goal with an aperture array. While each of the MWA SETI publications follows a similar processing and search approach, our data analysis techniques have been gradually and significantly improved. The observations toward Orion represented an improvement in imaging techniques and source finding. In the observations toward Vela, the data were collected with an updated ``Phase II" array, increasing the spatial resolution by more than a third (3\,arc\,minutes down to 1\,arc\,minute).  

\begin{figure*}
\begin{center}
\includegraphics[width=0.98\textwidth]{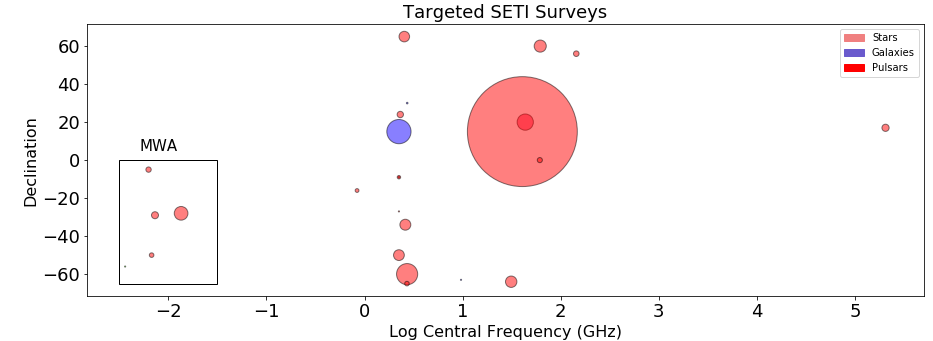}
\caption{A plot showing all targeted SETI searches published to date, as listed in https://technosearch.seti.org/radio-list and \cite{Enriquez_2017}. The x-axis is the central frequency of the survey and the y-axis represents the median declination the survey covered.  The colour of the circle represents the type of objects observed in the SETI survey and the size of the bubble is representative of the number of sources targetted in the survey.  This does not cover blind searches toward stars with no known exoplanets.  }\label{Bubbles}
\end{center}
\end{figure*}

In addition to the large field-of-view offered by the MWA, our surveys also represent the first published low radio frequency searches (see Figure \ref{Bubbles}). Since we don't know what frequency another technologically advanced civilisation might broadcast or operate at, there is no reason to ignore available search space. There are additional motivations for low-frequency as well. \cite{Sullivan_1978} suggested that the FM radio broadcasting stations of the world represents the greatest power per hertz in the radio band and this was further explored by \cite{Loeb_2007}. Overall,  there is growing support for broadening the frequency range searched for technosignatures to lower frequencies.  For example, \citet{Houston_2021} find that, according to several detection optimisation metrics, SETI surveys should be undertaken down into the hundreds of megahertz frequency range.

In this survey we utilise the procedures developed in our search toward Vela to search 200\,deg$^{2}$ toward the centre of our Galaxy but at a higher frequency of 155\,MHz.  This survey maintains the higher spatial resolution we obtained toward Vela, but encompasses the largest population of known exoplanets than across our previous surveys. 



\subsection{The Galactic Centre}

\begin{table*}
\caption{Parameters of previous MWA SETI surveys.} 
\centering
\begin{tabular}{lccccccc}
 & Phase centre & Phase center & Freq.  & FoV           &$\rm{RMS_{min}}$ & $\rm{EIRP_{min}}$ &Exoplanets  \\
 & (J2000)    &l,b (deg)   & (MHz)      & (deg${^2}$) & (Jy/beam) &  10$^{13}$(W) &  Known \\
\hline
\hline
\textbf{Galactic Centre }      & Phase I MWA  &&   &  & &&\\
Tingay et al. 2013      & 17h45m40s &0, 0 &103--133  & 400 & 0.45&$<$4 &38\\
    & $-$29d00m28s  &    &  & & &\\
\textbf{Orion}       & Phase I MWA & &    &  & & & \\
Tingay et al. 2018       & 05h35m17s &196, --15 &99--122     & 625 &0.28 & $<$1&22 \\
      & $-$05d23m28s &     & & & & &\\
\textbf{Vela} &  Phase I MWA &   & &  & & & \\
Tremblay \& & 08h35m27s& 264, --5 & 98--128 &  400 &0.034 & $<$0.6&6 \\
Tingay 2020&  $-$45d12m19s &  &  &  & & &\\
\textbf{Galactic Centre }      & Phase II MWA  && &  &  & &\\
This work                 & 17h45m40s&0, 0  & 139--169&   200&0.14  & $<$ 27& 144\\
     &$-$29d00m28s  & &  &  & &&\\
\hline
\end{tabular}

\end{table*}

The Galactic Centre (GC) is a prime SETI target as the line of sight toward the GC has the largest integrated count of Galactic stars for any direction. Here, we outline some arguments favouring and disfavouring the GC as a region where intelligent life may reside; however, we note that direct observational evidence remains the only method capable of proving the existence of life beyond Earth.

The high density of stars within the GC means that cataclysmic events such as stellar supernovae and magnetar flares are more likely to impact exoplanets within the GC, potentially destroying any life on their surface. Based on these factors, \citet{Lineweaver:2004} identify a `Galactic habitable zone' as an annular region between 7--9 kpc from the GC. Interactions due to close stellar flybys---more common within the dense GC---are also expected to damage planetary disks \citep{Lineweaver:2004, Jimenez-Torres:2013}; however, on longer timescales this could be advantageous to habitability.  Despite these factors, modelling by \citet{Gowanlock:2011} finds a majority of planets that may support complex life are found toward the inner Galaxy (less than 1\,kpc from the Galactic Centre). \citet{Morrison:2015} extend this model to include intelligent life, and also find higher probability within the inner Galaxy.  \citet{Gajjar:2021} update the model of \citet{Gowanlock:2011} to include the galactic bulge within the radius R $< 2.5$\,kpc, again finding the fraction of stars with a habitable planet is greatest in the inner regions of the Galaxy. Modelling by \citet{Cai:2021}, which includes factors such as abiogenesis (the idea that life arose from nonlife), evolutionary timescales, and  self-annihilation also find {a higher likelihood of intelligent life emerging in the Galactic inner disk (defined here as R $\leq$ 8 kpc), with peak likelihood at an annulus 4 kpc from the GC}.

Separate to the propensity of life to emerge, the GC's high stellar density may be advantageous to the growth of advanced space-faring societies. Such societies are likely to be capable of producing technosignatures detectable across large distances. First discussed in \citet{Newman:1981}, diffusion of advanced societies across the Milky way is accelerated in areas of high stellar density. \citet{DiStefano:2016} suggest that the close proximity of stars within dense globular clusters is favourable to the diffusion of space-faring societies. Modelling by  \citet{Carroll-Nellenback:2019} also suggests that high stellar density provides a front for settlements to rapidly expand from. Nevertheless, dust grains, gas and cosmic rays all pose hazards to interstellar travels, making the GC harder to traverse \citep{Lacki:2021}. The GC can also be considered a natural cynosure, or `Schelling point' within the Milky Way: an optimal location to place a transmitter to maximise the chances of its detection  \citep{Gajjar:2021}. 

Despite arguments that motivate SETI searches of the GC, only a handful of observations have been published. In 1981, the Westerbork Synthesis Radio Telescope was used to search for narrow band pulsating beacons in the GC \citep{Shostak:1985}. 
Our previous observations \citep[Paper 1,][]{Tingay_2016} searched toward 38 known exoplanets at a 10\,kHz resolution in 2 hours of data. As part of the Breakthrough Listen search for intelligent life, the GC is being surveyed using the Robert C. Byrd Green Bank Telescope
(GBT) and Parkes 64\,m telescope to cover 0.7--93GHz \citep{Worden:2017, Gajjar:2021}. Our observations, which are at lower frequency, complement the Breakthrough Listen search. Both Parkes and Green Bank are large single-dish telescopes with wide-bandwidth receivers. Interferometers, such as the MWA, offer wider fields of view and much better spatial resolution, giving improved signal localisation and radio frequency interference (RFI) rejection. However, compared to single-dish telescopes, interferometers have higher ingest data rates and require more demanding signal processing systems; consequently, their instantaneous bandwidth and spectral resolution are generally more constrained than single-dish instruments. As such, single-dish and interferometric approaches to technosignatures are complementary.

\begin{figure*}
    \centering
    \includegraphics[width=0.485\textwidth]{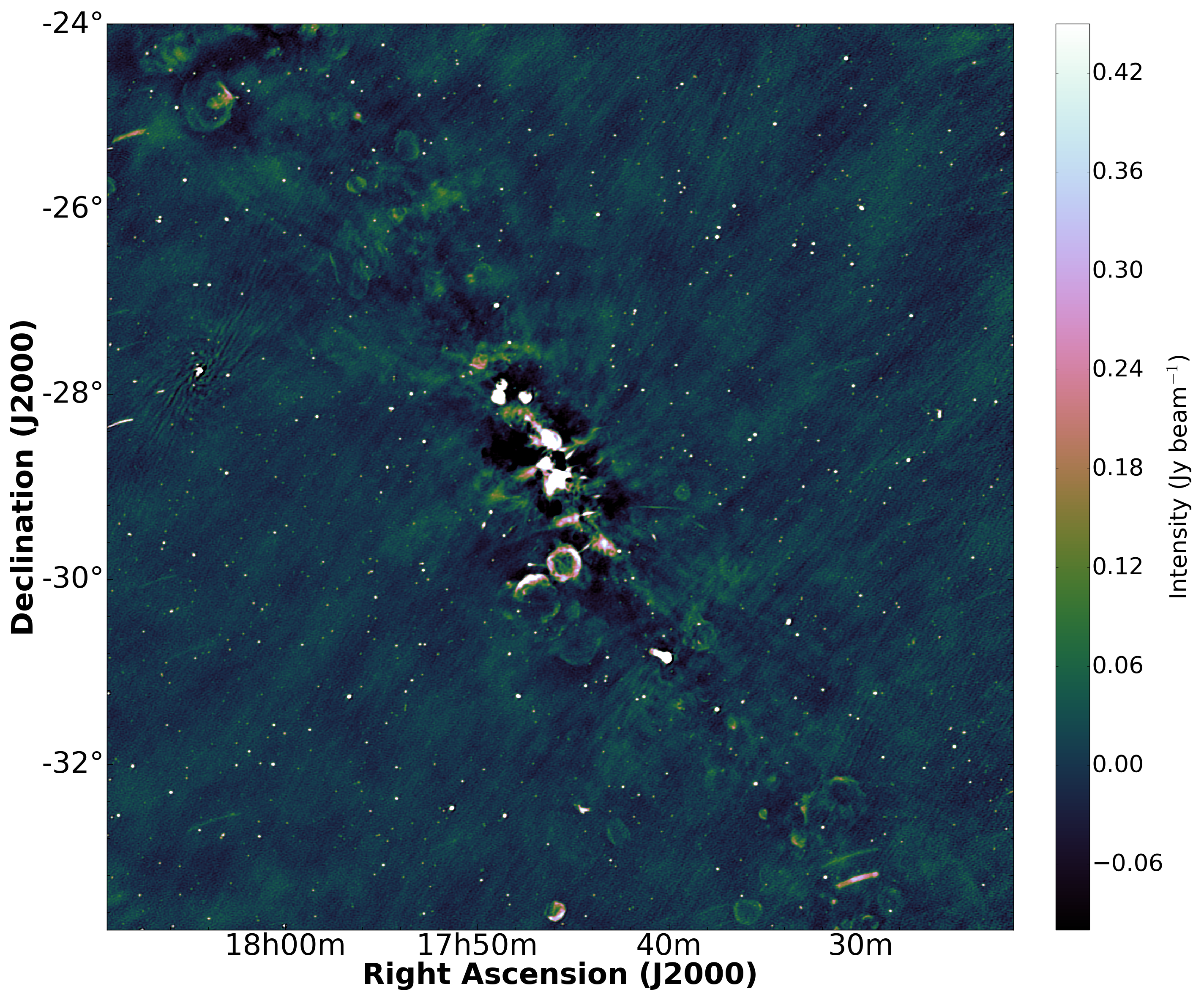}
    \includegraphics[width=0.48\textwidth]{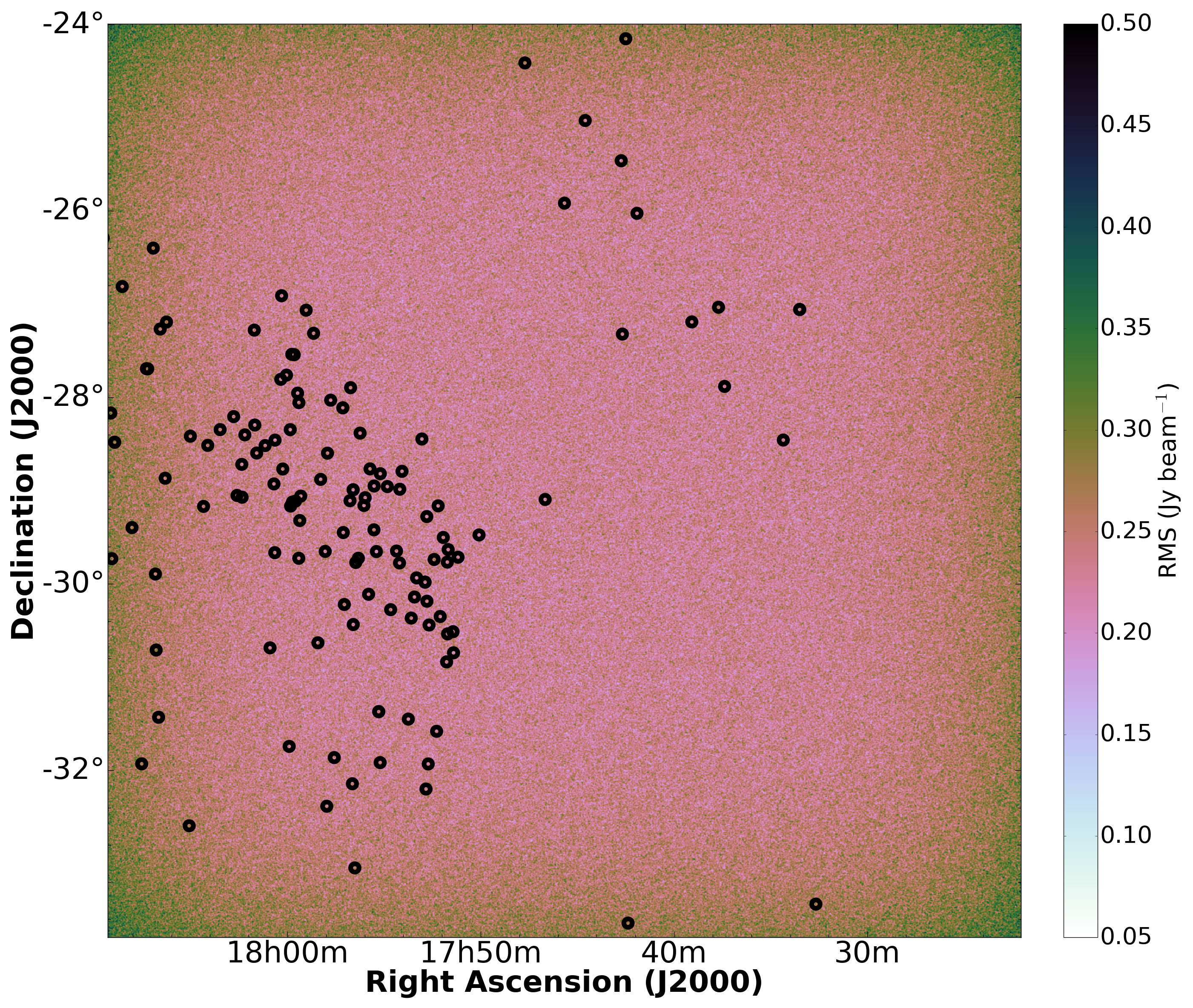}
    \caption{Continuum image of the Galactic Plane in ICRS Coordinates as viewed by the MWA at 155\,MHz (left).  The right-hand image is an image of the spectral RMS across the field, representative of the values extracted for the EIRP$_{min}$ limits. The black circles are the positions of the known sources in The Extrasolar Planet Catalogue with solar mass less than 60\,M$_{J}$ (a limit set by the catalogue custodians), as summarised in Tables \ref{tab3}, \ref{tab4},  \ref{tab5}, and \ref{tab6}.  }
    \label{Fig1}
\end{figure*}

As shown in Figure \ref{Fig1} (Section 3.1), the known exoplanets in this survey are distributed along either side of the Galactic plane (plus and minus longitude centred along b=0).  Microlensing is the most common method of detection and, as such, this population represents a population of planets significantly more distant than some of our previous surveys.

\begin{figure}
    \centering
    \includegraphics[width=0.48\textwidth]{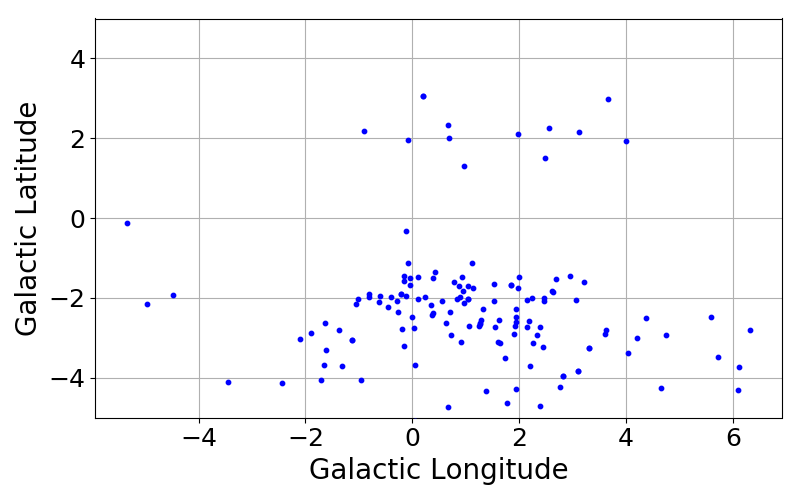}
    \caption{A plot of the sources listed in Tables \ref{tab3}, \ref{tab4},  \ref{tab5}, and \ref{tab6} shown in Galactic Coordinates. As shown there are more sources distributed along the negative latitude which is likely a selection effect from the exoplanet survey axis fields.}  
    \label{Gal_sources}
\end{figure}

\section{Observations}
The Murchison Widefield Array (MWA; \citealt{Tingay_2013}) is a low-frequency interferometer operating between 70 and 300\,MHz at the Murchison Radio-astronomy Observatory in Western Australia.  In 2018, the telescope was upgraded to the ``Phase II'' array \citep{Wayth_PhaseII}, doubling the number of aperture array tiles from 128 to 256 and approximately doubling the maximum baseline from 3\,km to 5.5\,km, of which 128 tiles are correlated at one time. These observations use the extended ``long base-line array'' with baselines between $\sim$22\,m and 5.5\,km.

Observations of the Galactic Centre, centred on J2000 coordinates 17:45:40.04 $-$29:00:28, were taken during 5--7 September 2020 for a total of 10.5\,hours and are summarised in Table\,\ref{obs}. The observations were taken when the Galactic Centre was low on the horizon\footnote{The MWA observations are obtained through competitive time allocation and when the observations happen are based on scientific priorities and work around maintenance schedules. Therefore preferential observational time when the Galactic Centre was closer to zenith was not available.}, so only 7 hours of the most sensitive observations were processed.  The data were calibrated from 2-minute observations of Hercules A each night and processed as described in \cite{Tremblay_2020_ApJ}.  To image these observations we used {\sc wsclean} version 2.9.2 with Briggs weighting of $-$1. Due to the large volume of these data, we did not process using other weightings.  This weighting improves our point-source sensitivity to reduce the potential effects of beam dilution over previous publications. These settings produced a cube with a field-of-view (FOV) of 200\,square\,degrees and a synthesised beam of 75$\times$67\,arc\,seconds.  Based on the bandpass solutions, three of the 128 tiles were flagged for every observation.

\begin{table}
\small
\caption{MWA Observing Parameters}
\label{obs}
\begin{tabular}{ll}
\hline
Parameter & Value\\
\hline
\hline
Central frequency& 154.4\,MHz\\
Total bandwidth & 30.72\,MHz\\
Number of imaged channels & 2400\\
Number Channels used in Search & 2280\\
Channel separation & 10 kHz \\
Synthesized beam FWHM$^{*}$ & 83$^{\prime\prime}$ $\times$ 67$^{\prime\prime}$\\
Imaged Region & 200\,sq. deg. \\
Phase centre of image (J2000) &  17h45m40s \\
&--29d00m28s\\
Total Integration Time & 7\,hours \\
\hline
$^*$Full Width at Half Maximum (FWHM)
\end{tabular}
\end{table}

The MWA uses a two-stage polyphase filter-bank which channelises the data into 24 $\times$ 1.28\,MHz ``coarse'' frequency channels which are then further divided into 128 $\times$ 10\,kHz ``fine'' frequency channels resulting in 3072 (10 kHz-wide) spectral channels. Each of the 85 5-minute observations were imaged independently using {\sc WSCLEAN} \citep{offringa-wsclean-2017} and a Stokes I primary beam model was created using the simulation-based Full Embedded Element (FEE) model of \cite{MWA_PB} for each of 24 coarse channels.  The creation of the model, taking into account the phase centre and other individual observational parameters in the metadata, was created using code by \cite{john_morgan_2021_5083990}. The primary beam model was applied to the 100 fine frequency channels imaged\footnote{Only 100 of the 128 fine frequency channels in each coarse channel were imaged to avoid channels were suffer from instrumental aliasing.} within each coarse channel to create primary beam corrected spectral channel images.  The images were then stacked together into a three-dimensional data cube for each observation, which is then time-averaged together using inverse variance weighting. After correction, the flux density error was determined by comparison to the Molonglo Reference Catalogue (MRC; \citealt{Large_1981}) which was scaled down to our frequency using a spectral index of $-$0.83, as described in detail in \cite{Hurley_Walker_2017}. The error across the field was 9$\pm$20\,mJy.

\begin{figure*}
    \centering
    \includegraphics[width=0.98\textwidth]{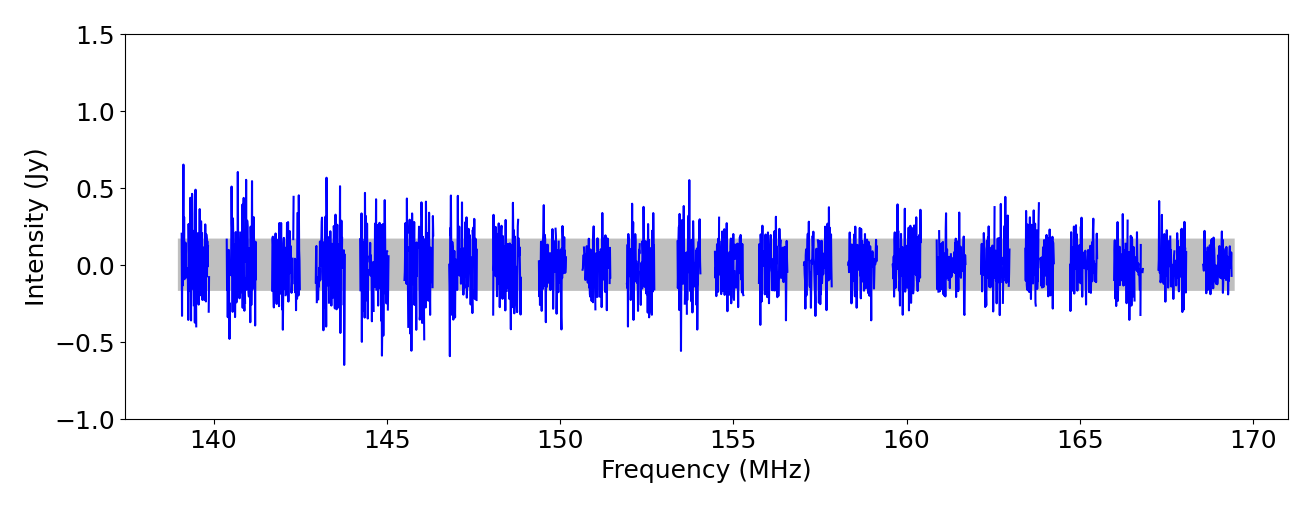}
  \caption{MWA spectrum, with a total integration time of 7 hours, at the position of KMT-2018-BLG-1292L b. Flagged channels and the edge channels which suffer from instrumental aliasing (see Section 2), are blanked out in the spectrum. The horizontal grey shaded region represents the $\pm1\sigma$ RMS value used in Table \ref{tab3}. }\label{Spectra}
\end{figure*}

Radio Frequency Interference (RFI) was flagged in two stages in the $u,v,w$ (visibility) domain.  First, each 5-minute observation was flagged using a statistical technique invoked by {\sc aoflagger} \citep{OffriingaRFI} and is a standard option for data downloaded through the All-Sky Virtual Observatory.  Secondly, a baseline flagging algorithm which flagged amplitudes in $u,v,w$ of baselines that were higher than three times the standard deviation\footnote{\url{https://gitlab.com/Sunmish/piip/blob/master/ms_flag_by_uvdist.py}}.  This flagged $<$1--3\% of the baselines. In previous work \citep{Tremblay_2020_ApJ}, strong FM band interference required a third-stage image-based filtering to flag channels where RFI was detected in a large fraction of pixels; this third-stage flagging was not required for these data. An example spectrum is shown in Figure \ref{Spectra}.

The ionosphere can create an astrometric shift in the sources which is more pronounced at lower frequencies. The estimated source position error is estimated in the final composite image by comparing to the MRC.  After correction, the systematic spatial error in a fully integrated continuum image was $+$5$\pm$8\,arcseconds in right ascension and $+$22$\pm$5\,arcseconds in declination. The ratio of point source flux density and peak intensity is 1.1$\pm$0.2.  This suggests that the ionospheric distortions are corrected for in these observations.

\section{Results}
\subsection{Known Exoplanets}
The three-dimensional data cubes were each searched in frequency for 10\,kHz (frequency resolution of the MWA) signals of non-astrophysical origin toward the 144 known exoplanets in our field.  This list of exoplanets from the The Exoplanet Encyclopaedia\footnote{\url{http://exoplanet.eu/}}, represents an accumulated list of exoplanets with a mass limit of 60\,M$_{J}$. In Tables \ref{tab3}, \ref{tab4}, and \ref{tab5} we provide information on the planet designation, position, distance, planet mass, detection method, sensitivity limit as spectral variation in our data cube toward that position, and the equivalent isotropic power limit (EIRP). In Table \ref{tab6} we provide the information of exoplanets in our sample which have no known distance, and therefore a limit on the EIRP could not be obtained.

Similar to our survey toward Vela \citep{Tremblay_Vela_SETI}, we calculated the EIRP upper limit from the equation:
\begin{equation}
    \mathrm{EIRP (W)} < 1.12\times10^{12} S_{\mathrm{rms}} R^{2} ,
\end{equation}
where S$_{\mathrm{rms}}$ is the RMS intensity value in Jy\,beam$^{-1}$ along the spectral frequency axis toward the source position and $R$ is the distance to the stellar system in pc.  Similar to our previous work, we assume a signal width of 10\,kHz to match the frequency resolution of the MWA, and discuss in Section 4.1 about potential for spectral broadening when higher resolution is assumed. In these upper limits we assume that the signal optimally fills the 10\,kHz channel width. If an artificial signal was an unresolved signal with a width between 1--10\,kHz but constrained within the channel, the estimated EIRP would be underestimated by 10kHz/$\Delta\nu$t, where $\Delta\nu$t is the transmission bandwidth. We do not account for additional edge cases where a signal spans the channel gap between two fine frequency channels. We note that for the EIRP Limit toward the known sources we use a one sigma value for direct comparison with previous work and a 6 sigma limit for the blind search. 

\begin{figure}
\centering
\includegraphics[width=0.49\textwidth]{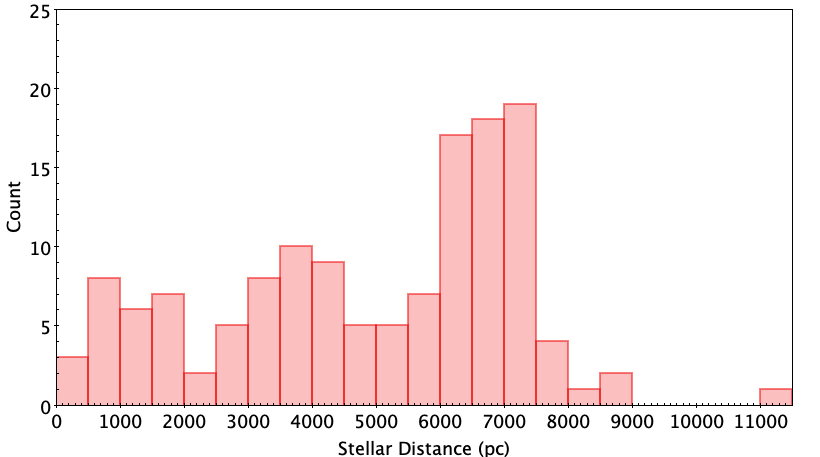}
\includegraphics[width=0.49\textwidth]{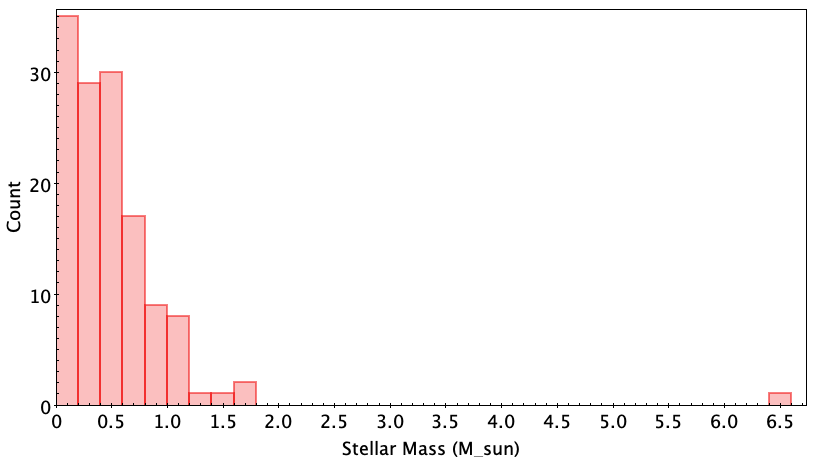}
\caption{Summary of the stellar parameters of the known sources in the exoplanet catalogue. As shown most exoplanets are around a solar mass or less and the distances are distributed up to 9\,kpc.}
\label{Stats}
\end{figure}

As shown in Figure \ref{Stats}, the majority of the stars containing exoplanets are near or below solar mass.  It is also shown that the sample of known exoplanets have a mean distance of 4.8\,kpc, which is much farther than our sample of stellar objects we studied toward Vela or Orion.

\begin{table*}
\caption{Known exoplanets in the survey field, from the exoplanet catalog: http://exoplanet.eu/} 
\centering
\begin{tabular*}{\textwidth}{@{}l\x c\x c\x c\x c\x c\x c\x c\x c@{}}
\hline \hline

Designation   & RA (J2000)   & Dec (J2000)     &  Distance  &  MSin(i) 
         & Period  & Detection  &  1$\sigma$ RMS    & EIRP$^{c}$    \\
         & (deg.)        & (deg.)          & (pc)                  & (M$_{J}$$^{a}$) 
         & (days) &  method$^{b}$             & (Jy/beam)                  & $<$10$^{13}$ (W)        \\
\hline
HD 164604 b	&	270.779	&	--28.561	&	39.4	&		&	641.47	&	RV	&	0.16	&	27.3	\\
HD 165155 b	&	271.488	&	--29.917	&	64.98	&		&	434.50	&	RV	&	0.16	&	77.9	\\
OGLE-2009-BLG-151L b	&	268.592	&	--29.056	&	390	&	7.50	&		&	Micro	&	0.14	&	2454	\\
OGLE-2015-BLG-0954L b	&	270.183	&	--28.661	&	600	&	4.40	&		&	Micro	&	0.15	&	6029.8	\\
MOA-2007-BLG-192L b	&	272.017	&	--27.150	&	700	&	0.01	&		&	Micro	&	0.18	&	10127.2	\\
OGLE-2018-BLG-0532 b	&	269.983	&	--28.998	&	790	&	0.02	&		&	Micro	&	0.14	&	10117.6	\\
MOA-2011-BLG-274 b	&	268.675	&	--28.916	&	800	&	0.80	&		&	Micro	&	0.15	&	10679.2	\\
MOA-2010-BLG-328L b	&	269.496	&	--30.715	&	810	&	0.03	&		&	Micro	&	0.15	&	10922.4	\\
OGLE-2017-BLG-1434L b	&	268.280	&	--30.246	&	860	&	0.01	&		&	Micro	&	0.15	&	12252.9	\\
OGLE-2019-BLG-0960 b	&	274.013	&	--25.773	&	880	&	0.01	&		&	Micro	&	0.16	&	13601.5	\\
OGLE-2013-BLG-0341L b	&	268.029	&	--29.846	&	911	&	0.01	&		&	Micro	&	0.14	&	13511.1	\\
OGLE-2017-BLG-1522L b	&	270.321	&	--28.462	&	990	&	0.75	&		&	Micro	&	0.15	&	16405.6	\\
OGLE-2013-BLG-0578L b	&	269.999	&	--29.735	&	1160	&	34.00	&		&	Micro	&	0.15	&	22762.2	\\
OGLE-2016-BLG-1227 b	&	265.596	&	--33.760	&	1210	&	0.79	&		&	Micro	&	0.18	&	29124.6	\\
KMT-2020-BLG-0414 b	&	271.917	&	--28.485	&	1220	&	0.00	&		&	Micro	&	0.18	&	29895.4	\\
KMT-2020-BLG-0414 c	&	271.917	&	--28.485	&	1220	&	23.30	&		&	Micro	&	0.18	&	29895.4	\\
OGLE-2014-BLG-0257L b	&	270.450	&	--28.262	&	1250	&	36.00	&		&	Micro	&	0.15	&	26529.6	\\
OGLE-2012-BLG-0563L b	&	271.491	&	--27.712	&	1300	&	0.39	&		&	Micro	&	0.17	&	32042.1	\\
OGLE-TR-10 b	&	267.867	&	--29.876	&	1500	&	0.68	&	3.10	&	PT	&	0.15	&	37511	\\
OGLE--TR-56 b	&	269.146	&	--29.539	&	1500	&	1.30	&	1.21	&	PT	&	0.14	&	36541.1	\\
OGLE-2006-109L b	&	268.146	&	--30.088	&	1510	&	0.73	&	1790.00	&	Micro	&	0.15	&	38090.3	\\
OGLE-2006-109L c	&	268.146	&	--30.088	&	1510	&	0.27	&	4931.00	&	Micro	&	0.15	&	38090.3	\\
OGLE-2014-BLG-1186L b	&	265.500	&	--34.288	&	1700	&	0.14	&		&	Micro	&	0.17	&	55313.1	\\
OGLE-2012-BLG-0358L b	&	265.696	&	--24.261	&	1760	&	1.85	&		&	Micro	&	0.17	&	60509.2	\\
OGLE-2011-BLG-0420L b	&	267.733	&	--29.825	&	1990	&	9.40	&		&	Micro	&	0.15	&	66967.7	\\
OGLE-2014-BLG-0676L b	&	268.104	&	--30.548	&	2220	&	3.09	&		&	Micro	&	0.15	&	82461.9	\\
MOA-2010-BLG-477L b	&	271.529	&	--31.454	&	2300	&	1.50	&		&	Micro	&	0.16	&	95860.4	\\
OGLE-2018-BLG-1269L b	&	269.692	&	--27.618	&	2560	&	0.69	&		&	Micro	&	0.15	&	109438	\\
KMT-2018-BLG-0029 b	&	264.471	&	--27.985	&	2730	&	0.02	&		&	Micro	&	0.15	&	127423	\\
OGLE-2007-BLG-349L (AB) b	&	271.350	&	--26.422	&	2760	&	0.25	&		&	Micro	&	0.17	&	143323	\\
OGLE-2015-BLG-0966L b	&	268.754	&	--29.047	&	2900	&	0.07	&		&	Micro	&	0.15	&	138517	\\
KMT-2018-BLG-1292L b	&	263.179	&	--33.521	&	2920	&	4.90	&		&	Micro	&	0.17	&	163666	\\
OGLE-2012-BLG-0950L b	&	272.019	&	--29.732	&	3000	&	0.11	&		&	Micro	&	0.18	&	184790	\\
OGLE-2013-BLG-0102L b	&	268.030	&	--31.691	&	3020	&	13.60	&		&	Micro	&	0.15	&	152114	\\
OGLE-2016-BLG-1266L b	&	267.854	&	--29.742	&	3050	&	11.90	&		&	Micro	&	0.15	&	156131	\\
OGLE-2013-BLG-0911L b	&	268.883	&	--29.254	&	3220	&	10.08	&		&	Micro	&	0.15	&	172151	\\
KMT-2019-BLG-0842 b	&	268.458	&	--29.878	&	3320	&	10.28	&		&	Micro	&	0.15	&	184695	\\
MOA-2010-BLG-117 b	&	271.957	&	--25.345	&	3400	&	0.51	&		&	Micro	&	0.19	&	249325	\\
OGLE-2016-BLG-0613L (AB) b	&	269.263	&	--28.116	&	3410	&	4.18	&		&	Micro	&	0.15	&	193939	\\
MOA-2013-BLG-605L b	&	269.679	&	--29.398	&	3600	&	0.07	&		&	Micro	&	0.15	&	221716	\\
MOA-2008-BLG-379L b	&	269.704	&	--29.803	&	3600	&	4.10	&		&	Micro	&	0.15	&	215639	\\
OGLE-2014-BLG-0124L b	&	270.621	&	--28.396	&	3600	&	0.64	&		&	Micro	&	0.15	&	221489	\\
OGLE-2019-BLG-0954L b	&	267.913	&	--29.611	&	3630	&	14.20	&		&	Micro	&	0.15	&	223793	\\
OGLE-2016-BLG-1067L b	&	273.205	&	--27.013	&	3730	&	0.43	&		&	Micro	&	0.16	&	246169	\\
OGLE-2017-BLG-1375 b	&	269.154	&	--30.311	&	3790	&	10.33	&		&	Micro	&	0.15	&	236189	\\
KMT-2019-BLG-1715 b	&	270.371	&	--28.777	&	3860	&	2.56	&		&	Micro	&	0.15	&	252963	\\
OGLE-2013-BLG-0132L b	&	269.765	&	--28.421	&	3900	&	0.29	&		&	Micro	&	0.15	&	262897	\\
OGLE-2016-BLG-1195L b	&	268.850	&	--30.207	&	3910	&	0.00	&		&	Micro	&	0.15	&	251866	\\

\hline \hline
\end{tabular*}\label{tab3}

\medskip
\tabnote{$^a$Mass of planet times the sine of orbit inclination, in Jupiter masses}
\tabnote{$^{b}$RV= Radial Velocity; I=Imaging; PT=Primary Transit; Micro= Microlensing}
\tabnote{$^{c}$Equivalent Isotropic Radiated Power using the specified 1$\sigma$ limit}
\end{table*}

\begin{table*}
\caption{Known exoplanets in the survey field, from the exoplanet catalog: http://exoplanet.eu/} 
\centering
\begin{tabular*}{\textwidth}{@{}l\x c\x c\x c\x c\x c\x c\x c\x c@{}}
\hline \hline

Designation   & RA (J2000)   & Dec (J2000)     &  Distance  &  MSin(i) 
         & Period  & Detection  &  1$\sigma$ RMS    & EIRP$^{c}$    \\
         & (deg.)        & (deg.)          & (pc)                  & (M$_{J}$$^{a}$) 
         & (days) &  method$^{b}$             & (Jy/beam)                  & $<$10$^{13}$ (W)        \\
\hline
OGLE-2017-BLG-0604 b	&	267.892	&	--30.947	&	3950	&	0.51	&		&	Micro	&	0.15	&	257703	\\
OGLE-2017-BLG-0896	&	264.879	&	--27.298	&	4000	&	19.00	&		&	Micro	&	0.15	&	264158	\\
OGLE-2006-BLG-284 b	&	269.658	&	--29.137	&	4000	&	0.45	&		&	Micro	&	0.15	&	271097	\\
OGLE-2012-BLG-0026L b	&	263.579	&	--27.143	&	4080	&	0.11	&		&	Micro	&	0.15	&	279574	\\
OGLE-2012-BLG-0026L c	&	263.579	&	--27.143	&	4080	&	0.68	&		&	Micro	&	0.15	&	279574	\\
OGLE-2011-BLG-0251L b	&	264.559	&	--27.136	&	4090	&	0.96	&		&	Micro	&	0.15	&	285529	\\
OGLE-2005-169L b	&	271.521	&	--30.733	&	4100	&	0.04	&		&	Micro	&	0.17	&	318741	\\
MOA-2007-BLG-197L b	&	271.771	&	--31.946	&	4170	&	41.00	&		&	Micro	&	0.18	&	344970	\\
OGLE-2015-BLG-1649 b	&	271.204	&	--32.633	&	4230	&	2.54	&		&	Micro	&	0.17	&	346455	\\
OGLE-2011-BLG-0265L b	&	269.450	&	--27.394	&	4380	&	0.88	&		&	Micro	&	0.15	&	327138	\\
OGLE-2016-BLG-1469L b	&	271.946	&	--26.290	&	4500	&	13.60	&		&	Micro	&	0.19	&	425740	\\
KMT-2017-BLG-0165L b	&	269.650	&	--28.134	&	4530	&	0.11	&		&	Micro	&	0.15	&	350648	\\
OGLE-2017-BLG-0173L b	&	267.971	&	--29.271	&	4700	&	9.08	&		&	Micro	&	0.14	&	358783	\\
OGLE-2014-BLG-1112L b	&	272.151	&	--28.666	&	4840	&	31.70	&		&	Micro	&	0.18	&	471270	\\
OGLE-2012-BLG-0406L b	&	268.325	&	--30.471	&	4970	&	2.73	&		&	Micro	&	0.15	&	405029	\\
OGLE-2015-BLG-1319L b	&	269.442	&	--32.472	&	5000	&	45.00	&		&	Micro	&	0.15	&	434763	\\
OGLE-2017-BLG-0406 b	&	269.000	&	--29.863	&	5200	&	0.41	&		&	Micro	&	0.15	&	446318	\\
OGLE-2003-BLG-235L b	&	271.317	&	--28.895	&	5200	&	2.60	&		&	Micro	&	0.16	&	481272	\\
MOA-2012-BLG-006 b	&	270.442	&	--29.109	&	5300	&	8.40	&		&	Micro	&	0.15	&	477922	\\
XTE J1807-294 b	&	271.752	&	--29.409	&	5500	&	14.50	&	0.03	&	Timing	&	0.17	&	585384	\\
OGLE-2017-BLG-1049 b	&	269.533	&	--27.144	&	5670	&	5.53	&		&	Micro	&	0.15	&	535062	\\
MOA-2009-BLG-387L b	&	268.463	&	--33.990	&	5700	&	2.60	&	1970.00	&	Micro	&	0.18	&	647709	\\
OGLE-2017-BLG-0482 b	&	269.049	&	--30.528	&	5800	&	9.00	&		&	Micro	&	0.15	&	568344	\\
OGLE-2018-BLG-0596 b	&	269.054	&	--29.199	&	5900	&	0.03	&		&	Micro	&	0.15	&	590643	\\
OGLE-2007-BLG-368L b	&	269.108	&	--32.238	&	5900	&	0.07	&		&	Micro	&	0.15	&	601071	\\
OGLE-2017-BLG-0373L b	&	269.329	&	--31.952	&	5900	&		&		&	Micro	&	0.15	&	599534	\\
KMT-2016-BLG-0212 b	&	268.439	&	--29.087	&	6000	&		&		&	Micro	&	0.15	&	597719	\\
OGLE-2014-BLG-1760L b	&	269.408	&	--28.963	&	6060	&	0.57	&		&	Micro	&	0.15	&	607185	\\
KMT-2017-BLG-1038 b	&	266.171	&	--25.143	&	6100	&	2.40	&		&	Micro	&	0.15	&	644335	\\
MOA-2009-BLG-319L b	&	271.742	&	--26.820	&	6100	&	0.21	&		&	Micro	&	0.17	&	729184	\\
OGLE-2011-BLG-0173L b	&	269.317	&	--28.684	&	6200	&	0.19	&		&	Micro	&	0.15	&	644380	\\
OGLE-2018-BLG-1428L b	&	265.550	&	--26.138	&	6220	&	0.77	&		&	Micro	&	0.15	&	661154	\\
KMT-2016-BLG-1820 b	&	268.767	&	--29.517	&	6260	&	4.57	&		&	Micro	&	0.15	&	657715	\\
OGLE-2018-BLG-0799 b	&	273.458	&	--25.486	&	6290	&	0.41	&		&	Micro	&	0.17	&	738527	\\
UKIRT-2017-BLG-001 b	&	266.654	&	--29.211	&	6300	&	1.28	&		&	Micro	&	0.15	&	653360	\\
OGLE-2013-BLG-1721L b	&	268.127	&	--30.293	&	6300	&	0.64	&		&	Micro	&	0.14	&	649382	\\
OGLE-2014-BLG-1722L b	&	268.754	&	--31.469	&	6400	&	0.17	&		&	Micro	&	0.15	&	674633	\\
OGLE-2014-BLG-1722L c	&	268.754	&	--31.469	&	6400	&	0.26	&		&	Micro	&	0.15	&	674633	\\
MOA-2016-BLG-227L b	&	271.475	&	--27.714	&	6400	&	2.30	&		&	Micro	&	0.17	&	778603	\\
KMT-2016-BLG-2605 b	&	269.825	&	--26.982	&	6421	&	0.77	&		&	Micro	&	0.15	&	696601	\\
MOA-2010-BLG-353L b	&	271.304	&	--27.293	&	6430	&	0.27	&		&	Micro	&	0.17	&	775990	\\
KMT-2016-BLG-2364 b	&	265.717	&	--27.436	&	6440	&	3.93	&		&	Micro	&	0.15	&	698233	\\
OGLE-2018-BLG-0962	&	268.175	&	--32.309	&	6470	&	1.37	&		&	Micro	&	0.15	&	727124	\\
OGLE-2005-390L b	&	268.579	&	--30.377	&	6500	&	0.02	&	3500.00	&	Micro	&	0.15	&	696976	\\
OGLE-2016-BLG-0263L b	&	269.895	&	--31.819	&	6500	&	4.10	&		&	Micro	&	0.15	&	722133	\\
KMT-2017-BLG-1146 b	&	269.104	&	--33.143	&	6600	&	0.85	&		&	Micro	&	0.16	&	773762	\\
KMT-2016-BLG-1397 b	&	272.667	&	--24.858	&	6600	&	7.00	&		&	Micro	&	0.17	&	860151	\\
OGLE-2012-BLG-0838 b	&	273.004	&	--25.712	&	6620	&	0.25	&		&	Micro	&	0.17	&	818145	\\

\hline \hline
\end{tabular*}\label{tab4}

\medskip
\tabnote{$^a$Mass of planet times the sine of orbit inclination, in Jupiter masses}
\tabnote{$^{b}$RV= Radial Velocity; I=Imaging; PT=Primary Transit; Micro= Microlensing}
\tabnote{$^{c}$Equivalent Isotropic Radiated Power using the specified 1$\sigma$ limit}
\end{table*}

\begin{table*}
\caption{Known exoplanets in the survey field, from the exoplanet catalog: http://exoplanet.eu/} 
\centering
\begin{tabular*}{\textwidth}{@{}l\x c\x c\x c\x c\x c\x c\x c\x c@{}}
\hline \hline

Designation   & RA (J2000)   & Dec (J2000)     &  Distance  &  MSin(i) 
         & Period  & Detection  &  1$\sigma$ RMS    & EIRP$^{c}$    \\
         & (deg.)        & (deg.)          & (pc)                  & (M$_{J}$$^{a}$) 
         & (days) &  method$^{b}$             & (Jy/beam)                  & $<$10$^{13}$ (W)        \\
\hline
KMT-2016-BLG-1107 b	&	266.417	&	--26.032	&	6651	&	3.28	&		&	Micro	&	0.15	&	750943	\\
OGLE-2015-BLG-1670L b	&	268.159	&	--28.552	&	6700	&	0.06	&		&	Micro	&	0.15	&	740142	\\
OGLE-2012-BLG-0724L b	&	268.967	&	--29.819	&	6700	&	0.47	&		&	Micro	&	0.15	&	743775	\\
KMT-2018-BLG-125L b	&	269.863	&	--27.878	&	6700	&	0.02	&		&	Micro	&	0.15	&	768189	\\
MOA-2013-BLG-220L b	&	270.988	&	--28.455	&	6720	&	2.74	&		&	Micro	&	0.16	&	798259	\\
OGLE-2016-BLG-1190L b	&	269.721	&	--27.614	&	6770	&	13.38	&	1223.60	&	Micro	&	0.15	&	778927	\\
MOA-2016-BLG-319L b	&	268.742	&	--29.751	&	6800	&	0.62	&		&	Micro	&	0.15	&	771666	\\
OGLE-2008-BLG-355L b	&	269.787	&	--29.241	&	6800	&	4.60	&		&	Micro	&	0.15	&	768759	\\
OGLE-2019-BLG-1053L b	&	270.167	&	--27.342	&	6800	&	0.01	&		&	Micro	&	0.15	&	783133	\\
MOA-2007-BLG-400L b	&	272.425	&	--29.224	&	6890	&	1.71	&		&	Micro	&	0.17	&	902075	\\
OGLE-2013-BLG-1761L b	&	268.408	&	--28.895	&	6900	&	2.80	&		&	Micro	&	0.15	&	787994	\\
MOA 2009-BLG-411L b	&	268.493	&	--29.749	&	6900	&	53.00	&		&	Micro	&	0.15	&	819836	\\
MOA-bin-29 b	&	269.375	&	--29.737	&	6900	&	0.55	&		&	Micro	&	0.14	&	773905	\\
MOA-2011-BLG-291 b	&	268.867	&	--29.171	&	7000	&	0.09	&		&	Micro	&	0.15	&	807552	\\
KMT-2016-BLG-2142 b	&	268.113	&	--29.384	&	7010	&	15.49	&		&	Micro	&	0.15	&	830574	\\
KMT-2019-BLG-0371 b	&	268.383	&	--31.555	&	7025	&	16.50	&		&	Micro	&	0.15	&	833111	\\
KMT-2019-BLG-1953L b	&	269.117	&	--28.201	&	7040	&	0.59	&		&	Micro	&	0.15	&	844163	\\
KMT-2019-BLG-1953L c	&	269.117	&	--28.201	&	7040	&	0.28	&		&	Micro	&	0.15	&	844163	\\
OGLE-2015-BLG-1771L b	&	268.800	&	--28.863	&	7070	&	0.43	&		&	Micro	&	0.15	&	823398	\\
OGLE-2018-BLG-0567L b	&	269.017	&	--27.987	&	7070	&	0.32	&		&	Micro	&	0.14	&	812416	\\
KMT-2016-BLG-1836L b	&	268.250	&	--30.041	&	7100	&	2.20	&		&	Micro	&	0.15	&	847778	\\
OGLE-2018BLG-1011L b	&	269.013	&	--29.083	&	7100	&	1.80	&		&	Micro	&	0.14	&	820430	\\
OGLE-2018BLG-1011L c	&	269.013	&	--29.083	&	7100	&	2.80	&		&	Micro	&	0.14	&	820430	\\
MOA-2015-BLG-337 a	&	271.949	&	--28.170	&	7100	&	9.80	&		&	Micro	&	0.18	&	1022880	\\
MOA-2015-BLG-337 b	&	271.949	&	--28.170	&	7100	&	0.11	&		&	Micro	&	0.18	&	1022880	\\
KMT-2019-BLG-1339	&	265.742	&	--25.574	&	7150	&	12.20	&		&	Micro	&	0.15	&	869033	\\
(OGLE-2019/BLG-1019 b)&&&&&&&&\\
MOA-2012-BLG-505L b	&	268.142	&	--32.040	&	7210	&	6.70	&		&	Micro	&	0.15	&	895489	\\
KMT-2018-BLG-0748 b	&	267.875	&	--30.646	&	7300	&	0.19	&		&	Micro	&	0.15	&	900185	\\
OGLE-2017-BLG-1140L b	&	266.883	&	--24.523	&	7350	&	1.63	&		&	Micro	&	0.17	&	1009590	\\
MOA-2011-BLG-262L b	&	270.096	&	--30.755	&	7350	&	0.06	&		&	Micro	&	0.15	&	899051	\\
MOA-2011-BLG-028L b	&	270.858	&	--29.213	&	7380	&	0.09	&		&	Micro	&	0.16	&	964267	\\
OGLE-2018-BLG-1185 b	&	269.792	&	--27.835	&	7400	&	0.03	&		&	Micro	&	0.15	&	912731	\\
OGLE-2018-BLG-1700L b	&	269.954	&	--28.529	&	7600	&	4.40	&		&	Micro	&	0.15	&	966195	\\
OGLE-2018-BLG-0677L b	&	268.750	&	--32.017	&	7700	&	0.01	&		&	Micro	&	0.15	&	1000070	\\
MOA-2011-BLG-293L b	&	268.913	&	--28.477	&	7700	&	4.80	&		&	Micro	&	0.14	&	966263	\\
MOA-2011-BLG-322L b	&	271.225	&	--27.221	&	7740	&	7.80	&		&	Micro	&	0.16	&	1107550	\\
OGLE-2015-BLG-0051L b	&	269.663	&	--28.032	&	8200	&	0.72	&		&	Micro	&	0.14	&	1098900	\\
SWEEPS-04 b	&	269.725	&	--29.189	&	8500	&	3.80	&	4.20	&	PT	&	0.14	&	1171210	\\
SWEEPS-11 b	&	269.763	&	--29.198	&	8500	&	9.70	&	1.80	&	PT	&	0.15	&	1236990	\\
XTE J1751-305 b	&	267.806	&	--30.623	&	11000	&	27.00	&	0.03	&	Timing	&	0.15	&	2034430	\\

\hline \hline
\end{tabular*}\label{tab5}

\tabnote{$^a$Mass of planet times the sine of orbit inclination, in Jupiter masses}
\tabnote{$^{b}$RV= Radial Velocity; I=Imaging; PT=Primary Transit; Micro= Microlensing}
\tabnote{$^{c}$Equivalent Isotropic Radiated Power using the specified 1$\sigma$ limit}

\end{table*}

\begin{table*}
\caption{Known exoplanets in the survey field, from the exoplanet catalog with no known distance: http://exoplanet.eu/} 
\centering
\begin{tabular*}{\textwidth}{@{}l\x c\x c\x c\x c\x c\x c\x c\x c@{}}
\hline \hline

Designation   & RA (J2000)   & Dec (J2000)     &  Distance  &  MSin(i) 
         & Period  & Detection  &  1$\sigma$ RMS    & EIRP$^{c}$    \\
         & (deg.)        & (deg.)          & (pc)                  & (M$_{J}$$^{a}$) 
         & (days) &  method$^{b}$             & (Jy/beam)                  & $<$10$^{13}$ (W)        \\
\hline
KMT-2017-BLG-2820	&	263.743	&	--28.548	&		&		&		&	Micro	&	0.15	&	N/A	\\
KMT-2019-BLG-2073	&	267.471	&	--29.588	&		&	0.19	&		&	Micro	&	0.14	&	N/A	\\
OGLE-2016-BLG-596L b	&	267.803	&	--30.850	&		&	12.20	&		&	Micro	&	0.15	&	N/A	\\
OGLE-2017-BLG-0560	&	267.964	&	--30.459	&		&	1.90	&		&	Micro	&	0.15	&	N/A	\\
OGLE-2019-BLG-0551	&	269.870	&	--28.841	&		&	0.02	&		&	Micro	&	0.15	&	N/A	\\
OGLE-2012-BLG-1323	&	270.077	&	--28.584	&		&	0.01	&		&	Micro	&	0.15	&	N/A	\\
OGLE-2016-BLG-1540	&	270.196	&	--28.360	&		&		&		&	Micro	&	0.14	&	N/A	\\
OGLE-2016-BLG-1928	&	270.380	&	--29.130	&		&	0.00	&		&	Micro	&	0.15	&	N/A	\\

\hline \hline
\end{tabular*}\label{tab6}

\tabnote{$^a$Mass of planet times the sine of orbit inclination, in Jupiter masses}
\tabnote{$^{b}$RV= Radial Velocity; I=Imaging; PT=Primary Transit; Micro= Microlensing}
\tabnote{$^{c}$Equivalent Isotropic Radiated Power using the specified 1$\sigma$ limit}
\tabnote{N/A = Not Applicable as distances are unknown}
\end{table*}

\subsection{Blind Signal Search}
Each of the 2400 continuum-subtracted fine-channel (10\,kHz) images, each containing 2000$\times$2000 pixels, are independently searched using the source-finding software {\sc Aegean} \citep{Hancock_Aegean}. This is done using the function ``slice'', to set which channel in the cube is searched, and setting a ``seed clip'' value of 5, in order to search the image for pixels with a peak intensity value greater than 5\,\,$\sigma$ (where $\sigma$ is set from an input RMS image).  {\sc Aegean} works by fitting Gaussian's to the pixel data and applies a correction for the background\footnote{The background is defined by the 50$^{th}$ percentile of flux distribution in a zone 30 times the size of the synthesised beam.} to calculate the flux density for potential sources. This source-finding threshold has the goal of detecting all signals $>$6\,$\sigma$, which may not be pixel centred. Following this search, we found two signals over the $>$10$^{9}$ voxels searched. This is consistent with expected number of spurious signals within the data set. We therefore, do not detect any unknown emission sources. 

In \cite{Tremblay_Vela_SETI} we used the $Gaia$ catalogue to approximate the number of stars in the field toward Vela.  However, due to the high dust extinction toward the Galactic Centre and Galactic bulge, optical surveys have difficulty detecting sources.  Instead we use The GALACTICNUCLEUS Survey (GNS; \citealt{Nogueras-Lara_2018,Nogueras-Lara_2019}) which observes the J, K, and H infrared emission of 6000\,pc$^{2}$ (0.3\,deg$^{2}$) around the Galactic Centre, Galactic Bulge, and surrounding area. In their 49 independent pointings using the European Southern Observatory VLT HAWK instrument, they determined the precise photometric properties of 3.3$\times$10$^{6}$ stars with an assumed distance of 8178 $\pm$ 13$_{stat}$\,pc \citep{Gravity}.

The GNS infrared survey only covers 0.15\% of our field and accounts for in excess of 3.3\,million stars. So our blind search will likely cover billions of stars (and background galaxies).  Using the distance of 8.1\,kpc, we place a limit on the EIRP of putative transmitters within our observing band of $<$1.1$\times$10$^{19}$\,W.

\section{Discussion}
The median distance for the 144 known exoplanet systems toward the Galactic Centre is 5585\,pc with the shortest distance at 39\,pc. This is compared to the 28\,pc for the 6 exoplanets in \cite{Tremblay_Vela_SETI}, 50\,pc for the 22 exoplanets examined by \citet{Tingay_2018} and $\approx$2000\,pc for the 45 exoplanets examined by \citet{Tingay_2016}.  The EIRP upper limits in this paper represent some of the highest limits in our surveys to date, but as this is the first published survey in this frequency range we find this an important starting point. 

There are several metrics which allow for context of the transmitter values and the potential for detecting other societies. An extrapolation of terrestrial technology from the Kardashev \citep{1964SvA.....8..217K} scale, where a Type I civilisation with technology close to Earth is predicted to be able to emit an non-isotropically radiated signal at $\sim10^{17}$,W, is one suggested method.  We note that such a civilisation would only need to be able to build a big dish. \cite{Sullivan_1985} suggest that with an Arecibo-like single dish we could detect passive radio signals (or radio leakage) out to 9\,pc and with a large array of smaller dishes the distance could be more than 10 times that.  Overall, they concluded the most delectable signal from another planet would be a powerful military satellite.

Using examples on Earth, especially in the frequency range of the MWA, we can look at the Air Force Space Surveillance System known as ``Space Fence'' which operated up until 2013.  This system was a 1\,MW continuous wave (0.1\,Hz BW) system operating at 216\,MHz illuminating a 120 deg $\times$ 1.5\,arc\,min field at an EIRP of approximately 1.5$\times$10$^{10}$W \citep{Sullivan_1985}. If we compare this with our upper limits for the closest source, we are still three orders of magnitude from detecting a signal of this strength.

Similar to the metric used by \cite{Sheikh_2020}, we could measure our limits against the Arecibo Planetary Radar experiment ($\sim$20$\times$10$^{12}$\,W transmitted through a 305m parabolic reflector; L$_{A}$; \citealt{Siemion:2013}).  Their largest value for the EIRP limit for stars in 7--143\,pc is an upper limit in L$_{A}$ of 0.88. However, with our frequency resolution of 10\,kHz and the distances of 39\,pc or greater, the smallest L$_{A}$ upper limit is $\sim$13. 
This is significantly larger than our results from our other MWA surveys, which represented a population of closer stars.

Currently the MWA is undergoing an upgrade which will be pertinent to future SETI advances. Over the next 6 months, the MWA will be upgraded to increase frequency and time resolution, which will better match the kHz and sub-kHz searches completed by other facilities. An alternative approach to narrow-band SETI, first suggested by \citet{Cole:1979}, is to search for impulsive wide-band technosignatures; however comparatively few searches have been conducted. \citet{Gajjar:2021} search for artificial transient signals (0.7--194 ms duration) from the GC, using non-physical dispersion as a discriminant. Our interferometric data does not have sufficient time resolution for such a search, and at low frequencies signals are severely broadened on sight-lines (as discussed in Section 4.1) with high electron density, such as toward the GC. Nevertheless, the upgraded in higher time resolution modes which could be employed to search for impulsive low-frequency technosignatures toward the GC or other target fields. 

\subsection{Spectral broadening due to intervening media}

Electron density fluctuations in the interstellar medium (ISM), solar wind and interplanetary medium (IPM) cause spectral broadening by scattering signals as they are propagating through them. This is of particular concern for low-frequency observations of the GC, as the effect is stronger at low frequencies, and the large electron density fluctuations in the GC cause strong scattering. Although the MWA capabilities at the time of these observations have a relatively wide frequency resolution of 10\,kHz, future upgrades planned for 2021 will allow for sub-kHz resolution.  We therefore explore here the potential impact of future observations in this section.

\begin{figure*}
\begin{center}
\includegraphics[width=0.48\textwidth]{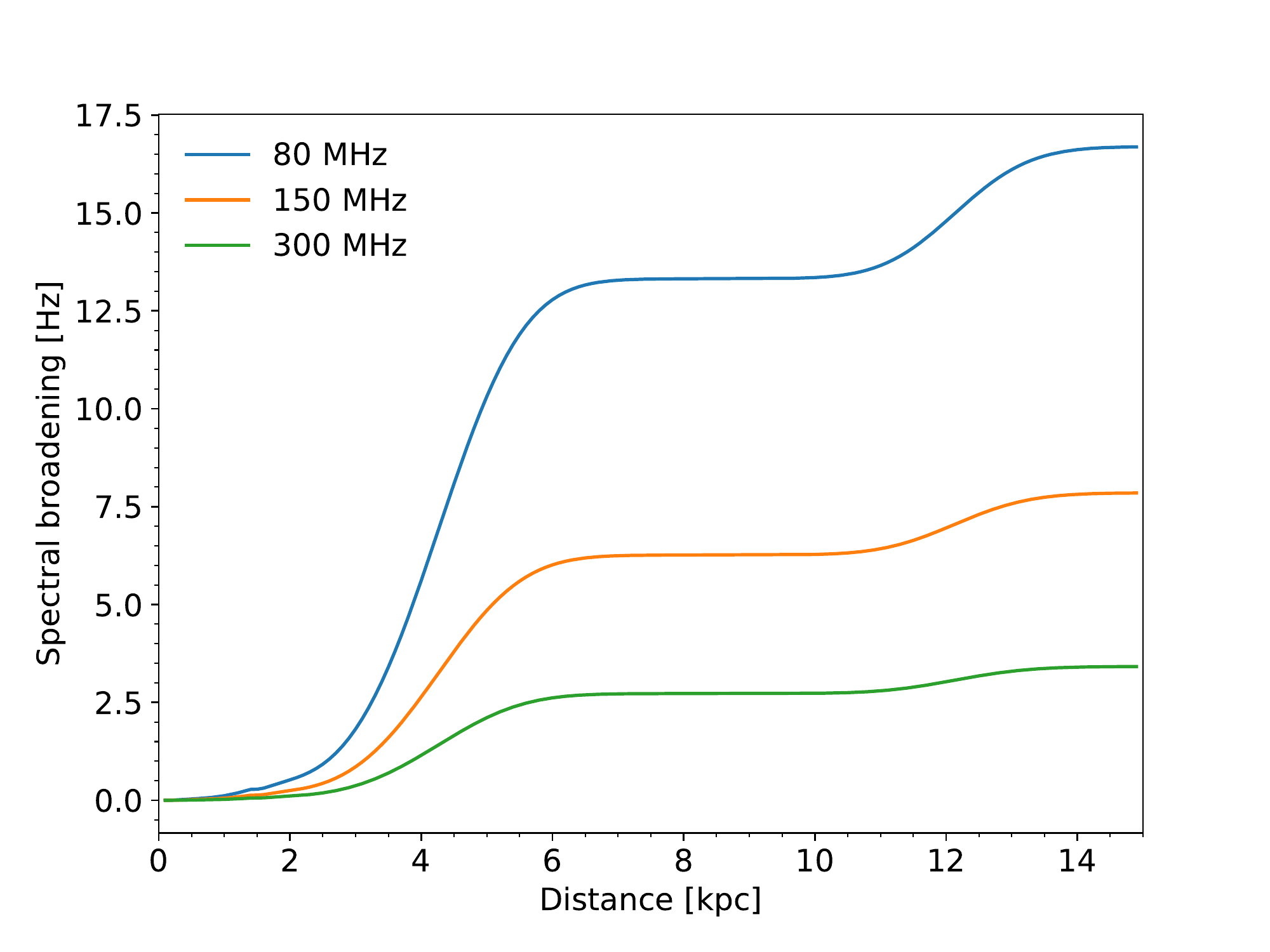}
\includegraphics[width=0.48\textwidth]{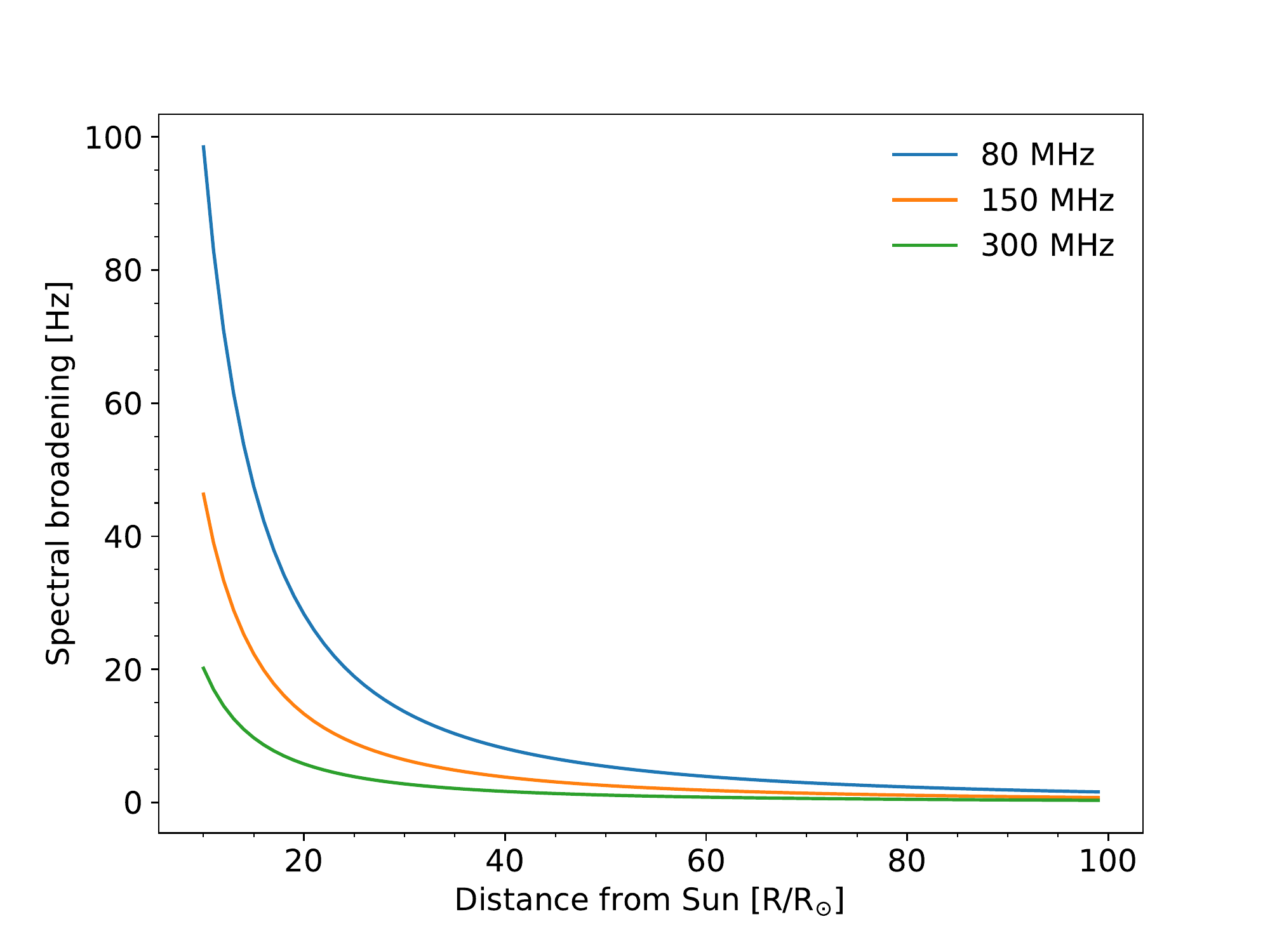}
\caption{Estimates of spectral broadening due to the interstellar medium (left) and interplanetary medium (right), for narrowband signals propagating from the direction of the GC.}\label{fig:spec-broad}
\end{center}
\end{figure*}

The effect of the ISM on narrowband signals is detailed in \citet{Cordes:1991} and \citet{Cordes:1997}. In the strong scattering regime a narrowband sinusoid with frequency $\nu_{\rm{GHz}}$ in GHz will be broadened by 
\begin{equation}
    \Delta\nu_{\rm{broad}}= 0.14\,{\rm{Hz}}~\nu_{\rm{GHz}}^{-6/5} \left(\frac{V_{\perp}}{100}\right) \rm{SM}^{3/5} , \label{eq:broad-ism}
\end{equation}
where $\Delta\nu_{\rm{broad}}$ is the spectral broadening in Hz, $V_{\perp}$ is the transverse velocity of the source in km s$^{-1}$, and SM is the scattering measure along the line of sight. The SM is an integrated measure of electron density fluctuations $C^2_{n_{e}}$ along the line of sight to a distance $L$, \citep{Rickett:1990}, defined as 
\begin{equation}
   {\rm{SM}} = \int_0^L C^2_{n_{e}}(z) dz.
\end{equation}
Estimates of SM toward the GC can be found using the NE2001 Galactic electron density model \citep{Cordes:2002}. To estimate the spectral broadening of narrowband signals at MWA frequencies, we use the \textsc{PyGEDM} code \citep{Price:2021} to generate SM from NE2001 as a function of distance toward the GC (Fig.\,\ref{fig:spec-broad}, left). To avoid a discontinuity at the GC ($l$=0, $b$=0), we add a 0.5 deg offset in Galactic latitude when calculating SM. {The highest radial velocities within the Galaxy occur toward the GC, reaching $|V_{\perp}|\approx$280\,km\,s$^{-1}$ \citep{Dame:2001}.} Applying Eq.\,\ref{eq:broad-ism},  we find spectral broadening of under {7.5 Hz} at an observational frequency of 150 MHz: well below our channel resolution of 10\,kHz.  

Following \citet{Siemion:2013}, spectral broadening due to the IPM can be estimated as
\begin{equation}
    \Delta\nu_{\rm{broad}}= 300\,{\rm{Hz}}~\nu_{{\rm{GHz}}}^{-6/5}\left(\frac{R}{R_{\odot}}\right)^{-9/5}\label{eq:broad-ipm}
\end{equation}
where $R_{\odot}$ is the radius of the Sun, and $R$ is the solar impact distance. The $R^{-9/5}$ dependence is detailed in \citet{Woo:2007}, based on an empirical fit to spectral broadening measured from the \emph{Pioneer} 10, 11 and \emph{Helios} 1 space mission telemetry, {and corroborated by radar reflections from Venus at superior conjunction \citep{Harmon:1983}}; this power law holds from $\sim10R_{\odot}$ to $\sim200R_{\odot}$. Solar activity and coronal mass ejections can nonetheless cause significant variations. Fig.\,\ref{fig:spec-broad} (right) shows the estimated effect of the IPM from Eq.\,\ref{eq:broad-ipm} for MWA frequencies. The effect of the IPM can be more significant than the ISM, particularly when the solar impact distance is small (i.e., the Sun passes close to the GC with respect to Earth).

Electron density fluctuations also impart scintillation on propagating signals, the characteristic timescale of which is given by
\begin{equation}
     \Delta t_{d} = 3.3\,s~\nu_{{\rm{GHz}}}^{-6/5}\left(\frac{V_{\perp}}{100}\right)^{-1} \rm{SM}^{-3/5}.
\end{equation}
If $\Delta t_d$ is much shorter than the observation length, scintillation effects will be quenched; however, if $\Delta t_d$ is much longer than the observation length, the signal may be effectively amplified or attenuated by scintillation during the observation, simplifying or confounding detection above our sensitivity threshold \citep{Cordes:1991}. Only the two nearest known exoplanets in the survey field, HD 164604 b and HD 165155 b, have $\Delta t_d$ estimates  significantly larger than our 7-hr observation length (using SM derived from NE2001). These sources could be re-observed at a later date, in case of scintillation-induced attenuation.

For our observations at 150 MHz, neither ISM nor IPM-induced broadening are significant effects. However, these effects should be considered in $\sim$Hz frequency resolution observations at low frequencies, in particular for proposed Moon-based SETI missions at frequencies below 1\,MHz \citep[e.g.][]{Michaud:2021} and future upgrades to the MWA. Signals from exoplanets that are occulted by their host star (with respect to Earth) can have large stellar impact distances and consequently large spectral broadening. The space weather around an exoplanet's host star will also affect broadening and scintillation; we note this as a potential field for future research.


\section{Conclusions}

We searched sightlines toward 144 known explanatory systems for artificial signals, finding no plausible technosignatures above an EIRP$_{\rm{min}}$ from 10$^{13}$\,W to 10$^{19}$\,W (depending on distance to the source). We also conducted a blind survey across a 200\,deg$^{2}$ field toward the GC, which covers billions of stars, including 3.3 million stars within 6000 pc$^2$ (0.3 deg$^2$) of the GC from the GALACTICNUCLEUS survey. Our observations across 139--170\,MHz are complementary to \citet{Tingay_2013} (103--133\,MHz) and higher-frequency (0.7--93\,GHz) technosignature searches currently underway with the Parkes and Green Bank telescopes \citep{Gajjar:2021}. Combined, these observations will allow statistical limits on the prevalence of putative transmitters at radio wavelengths in the GC. Future GC observations with the MWA and other instruments could allow full frequency coverage from 80\,MHz to 93\,GHz. Planned improvements to the MWA, due to be online in late 2021, will provide data with finer frequency and time resolution, which will improve the MWA's sensitivity to narrowband technosignatures and pulsating signals. Continual improvement of telescope capabilities, when combined with methodical observational campaigns, provides a means to explore the vast parameter space within which signs of technologically-capable life may be waiting to be found.

\section{Acknowledgements}

We thank A. Siemion and V. Gajjar for discussion on spectral broadening. We also thank the anonymous reviewers for their helpful comments and suggestions to improve the manuscript.

\subsection {Facilities}
This scientific work makes use of the Murchison Radio-astronomy Observatory, operated by CSIRO. We acknowledge the Wajarri Yamatji people as the traditional owners of the Observatory site. Support for the operation of the MWA is provided by the Australian Government (NCRIS), under a contract to Curtin University administered by Astronomy Australia Limited.  

\subsection{Computer Services}
We acknowledge the Pawsey Supercomputing Centre which is supported by the Western Australian and Australian Governments. Access to Pawsey Data Storage Services is governed by a Data Storage and Management Policy (DSMP). The All-Sky Virtual Observatory (ASVO) has received funding from the Australian Commonwealth Government through the National eResearch Collaboration Tools and Resources (NeCTAR) Project, the Australian National Data Service (ANDS), and the National Collaborative Research Infrastructure Strategy. This research has made use of NASA’s Astrophysics Data System Bibliographic Services.

\subsection{Software}
The following software was used in the creation of the data cubes:
\begin{itemize}
    \item {\sc aoflagger} and {\sc cotter} -- \cite{OffriingaRFI}
    \item {\sc WSClean} -- \cite{offringa-wsclean-2014,offringa-wsclean-2017}
    \item {\sc Aegean} -- \cite{Hancock_Aegean}
    \item {\sc miriad} -- \cite{Miriad}
    \item {\sc TOPCAT} -- \cite{Topcat}
    \item{NumPy v1.11.3 \citep{NumPy}, AstroPy v2.0.6 \citep{Astropy}, SciPy v0.17.0 \citep{SciPy}, Matplotlib v1.5.3 \citep{Matplotlib}}
    \item {\sc CARTA} -- \cite{angus_comrie_2020_3746095}
    \item {\sc PyGEDM} -- \citet{Price:2021}
\end{itemize}

\bibliographystyle{pasa-mnras}
\bibliography{mwa-seti.bib}

\end{document}